\documentstyle[epsfig]{aipproc}

\newcommand{\epsfigure}[2]{\epsfig{file=#1,width=#2}}

\begin{document}
\title{Absence of Replica Symmetry Breaking in a Region of the Phase Diagram
of the Ising Spin Glass}

\author{Hidetoshi Nishimori$^*$ and David Sherrington$^{\dagger}$}
\address{$^*$Department of Physics, Tokyo Institute of Technology\\
Oh-okayama, Meguro-ku, Tokyo 152-8551, Japan
\\
$^{\dagger}$Department of Physics, University of Oxford\\
Theoretical Physics, 1 Keble Road, Oxford OX1 3NP, UK}

\maketitle

\begin{abstract}
We prove that the distribution functions of magnetization and
spin glass order parameter coincide on the Nishimori line in the
phase diagram of the $\pm J$ Ising model in any dimension.
This implies absence of replica symmetry breaking because the
distribution function of magnetization consists only of two
delta functions, suggesting the same simple structure for the distribution
of spin glass order parameter.
It then follows that the mixed (glassy) phase, where the ferromagnetic
order coexists with complex phase space, should
lie, if any, below the Nishimori line.
We also argue that
the AT line to mark the onset of RSB
with a continous distribution of the spin glass order parameter,
if any again,
would start with an infinite slope from the multicritical point
where paramagnetic, ferromagnetic and spin glass phases merge.
\end{abstract}

\section*{Introduction}

Existence and characteristics of the spin glass phase
are actively investigated for finite-dimensional random spin systems.
Closely related is the problem of the mixed (glassy) ferromagnetic phase.
If the mean-field picture applies to finite-dimensional systems,
the ferromagnetic phase would split up into two regions,
one with a simple structure (the replica-symmetric (RS) phase in the
mean-field framework) and the other with a complex phase space
(replica-symmetry broken (RSB) state).
The boundary between these two phases would be the
finite-dimensional counterpart of the AT (de Almeida-Thouless) line
found for the infinite-range SK (Sherrington-Kirkpatrick) model.

Investigations of these and related problems are
almost exclusively carried out currently by numerical methods
because of lack of reliable analytical methods.
We show in the present contribution that simple symmetry arguments
lead to a strong constraint on the possible location
and shape of the AT line in the phase diagram of the Ising spin
glass on an arbitrary lattice with arbitrary range of interactions.
More precisely, it is possible to prove that there is nothing
like RSB on the Nishimori line in the phase diagram
and that the AT line, if it ever exists, should start as a vertical
line from the multicritical point where paramagnetic, ferromagnetic
and spin glass phases merge.
Our method is an application of the gauge theory which has been
used to derive exact energy and many other exact/rigorous results.

\section*{Gauge theory of the Ising spin glass}

Let us consider the $\pm J$ Ising model ($S_i=\pm 1$)
with the Hamiltonian
\begin{equation}
  H=-\sum_{\langle ij\rangle } J_{ij} S_i S_j,
   \label{nishimori:Hamiltonian}
\end{equation}
where the sum is over pairs of sites on an arbitrary
lattice with arbitrary range of interactions.
The exchange interactions are quenched random variables with the
distribution function
 \begin{equation}
   P(J_{ij}) = p\delta (J_{ij}-J)+ (1-p) \delta (J_{ij}+J).
     \label{nishimori:J-distr}
 \end{equation}
The Hamiltonian (\ref{nishimori:Hamiltonian}) is invariant under
gauge transformation
  \begin{equation}
     J_{ij}\to J_{ij}\sigma_i \sigma_j,~~~S_i\to S_i \sigma_i,
     \label{nishimori:g-trsf}
  \end{equation}
where $\sigma_i$ is an arbitrarily fixed Ising spin.

\begin{figure}[t] 
\begin{center}
\epsfigure{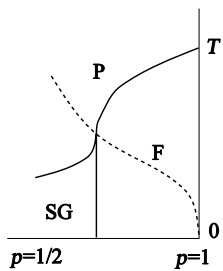}{1.5in}
\end{center}
\caption{Phase diagram of the $\pm J$ model and the Nishimori line (dashed).}
\label{nishimori:fig1}
\end{figure}

The gauge invariance of the Hamiltonian leads to a number of
exact and/or rigorous results \cite{nishimori:hn81,nishimori:hn2001}.
For example, consider the internal energy
 \begin{equation}
  E(K, K_p)= \left[ \langle H \rangle_K \right]_{K_p},
    \label{nishimori:E-def}
 \end{equation}
where the inner triangular brackets denote a thermodynamic average
and the outer square brackets represent the configurational average 
at a fixed value of $p$.
The suffixes denote the temperature and $p$ through
$K=\beta J=J/k_B T$ and $K_p=\frac{1}{2}\log p/(1-p)$.
The exact value of the internal energy can be calculated explicitly under
the constraint $K_p=K$ to give
 \begin{equation}
   E(K, K)=-N_B J \tanh K,
    \label{nishimori:exact-E}
  \end{equation}
where $N_B$ is the number of bonds on the lattice under consideration.
The constraint $K=K_p$ defines a curve in the phase diagram shown
dashed in Figure \ref{nishimori:fig1}, which is often called
the Nishimori line.

There are many other results including an upper bound on the specific heat
 \begin{equation}
   k_B T^2 C(K, K)\le \frac{J^2 N_B}{\cosh^2 K},
   \label{nishimori:c-bound}
 \end{equation}
a correlation identity
  \begin{equation}
    \left[ \langle S_i S_j \rangle_K \right]_K
      =\left[ \langle S_i S_j \rangle_K^2 \right]_K,
      \label{nishimori:corr-identity}
   \end{equation}
and a correlation inequality
  \begin{equation}
    \left[ \langle S_i S_j \rangle_K \right]_{K_p}
      \le 
    \left[ \left| \langle S_i S_j \rangle_{K_p} \right| \right]_{K_p}.
           \label{nishimori:corr-ineq}
   \end{equation}
The correlation identity (\ref{nishimori:corr-identity}) means that
the Nishimori line $(K_p=K)$ avoids the spin glass phase because
the left-hand side reduces to the square of the ferromagnetic
long range order $m^2$ as the distance between $i$ and $j$ tends to
infinity whereas the right-hand side approaches the square of the
spin glass order parameter $q^2$ in the same limit.
The spin glass phase is characterized by $q>0, m=0$ which is
inconsistent with the relation $m=\pm q$.
The correlation inequality (\ref{nishimori:corr-ineq}) places
a constraint on the possible shape of the boundary between the
ferromagnetic and spin glass phases:
This boundary can either be vertical or re-entrant in the
phase diagram.
The ferromagnetic phase is not allowed to lie below the spin
glass phase \cite{nishimori:hn81,nishimori:hn2001}.

\section*{Distribution of order parameters}

The gauge theory summarized above applies to derive an
identity between the distribution functions of the ferromagnetic
and spin glass order parameters on the Nishimori line
\cite{nishimori:hn2001,nishimori:SGN}.

\subsection*{Definitions and the main result}

The distribution function of magnetization is defined as
 \begin{equation}
   P_m(x; K)=\left[ \frac{\sum_S \delta (x-N^{-1} \sum_i S_i )
     e^{-\beta H(S)}}
     {\sum_S e^{-\beta H(S)}} \right]_{K_p}.
  \label{nishimori:Pm}
 \end{equation}
To define the distribution function of the spin glass order parameter,
it is convenient to introduce two replicas
of the same system with spins $\{S_i\}$ and $\{\sigma_i\}$ and
the couplings $K_1$ and $K_2$, respectively:
 \begin{equation}
   P_q(x; K_1, K_2)=\left[ \frac{\sum_S \sum_{\sigma}
   \delta (x-N^{-1} \sum_i S_i \sigma_i)
     e^{-\beta_1 H(S)}e^{-\beta_2 H(\sigma )}}
     {\sum_S \sum_{\sigma}
     e^{-\beta_1 H(S)}e^{-\beta_2 H(\sigma )}} 
     \right]_{K_p},
  \label{nishimori:Pq}
 \end{equation}
where $K_1=\beta_1 J$ and $K_2=\beta_2 J$.
The distribution of the spin glass order parameter is defined
in terms of two replicas with the same temperature, $P_q(x;K)
\equiv P_q(x; K, K)$.
Note that the distributions $P_m(x;K), P_q(x;K,K_p)$ and
$P_q(x;K)$ are all functions of $p$ as well.

\begin{figure}[t]
\begin{center}
\epsfigure{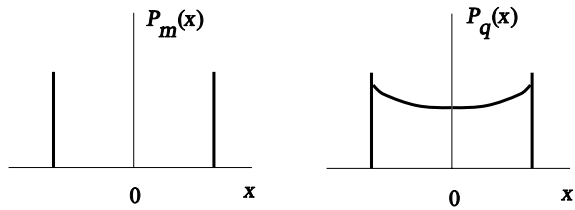}{3.3in}
\end{center}
\caption{Distribution of magnetization (left) and the SG order parameter (right).}
\label{nishimori:fig2}
\end{figure}

It is well established that the distribution of magnetization has
a simple structure with two delta functions as depicted in the left part
of Figure \ref{nishimori:fig2}.
The spin glass order parameter, on the other hand, has a more
complex structure.
For example, if there exists a full RSB,
the distribution has a continuous part as depicted schematically
in the right part of Figure
\ref{nishimori:fig2}.
Or, it may have delta peaks at a few additional locations
in the case of finite-step RSB.
In any case, the complexity of the system is closely
related with the structure of the distribution function
$P_q(x)$ different from that of $P_m(x)$.

Our main result is the following relation:
 \begin{equation}
   P_m(x;K)=P_q(x;K,K_p)
    \label{nishimori:pmq-relation1}
  \end{equation}
which holds for any combination of $x, K$ and $K_p$.
In particular, if we set $K_p=K$, Equation (\ref{nishimori:pmq-relation1})
reduces to the relation between the usual distribution functions of
magnetization and spin glass order parameter:
 \begin{equation}
   P_m(x;K_p)=P_q(x;K_p).
   \label{nishimori:pmq-relation2}
  \end{equation}
This immediately proves that the distribution function of the spin
glass order parameter on the right-hand side has a simple structure
represented by the magnetization distribution on the left-hand side if $K=K_p$.

The simplicity of the exact energy (\ref{nishimori:exact-E})
and a few other results \cite{nishimori:hn2001,nishimori:hn81}
suggest that the phase space of
the system would not be complicated.
The above result is regarded as definite evidence to support
this anticipation.

It also follows from differentiation of Equation
(\ref{nishimori:pmq-relation1}) that the derivatives of the
distribution functions $P_m(x;K)$ and
$P_q(x;K)$ with respect to $K$ and $p$ vanish when $K=K_p$:
  \begin{equation}
    \frac{\partial}{\partial K}P_q(x;K)=0,~~~
    \frac{\partial}{\partial p}P_q(x;K)=0
    \label{nishimori:pq-derivative}
  \end{equation}
for almost all $x$ when $K=K_p$.
These relations have been derived from the fact that the distribution
of magnetization $P_m(x;K)$ consists just of two delta functions
and therefore its derivatives with respect to the parameters $K$ and $p$
vanish for almost all $x$.
Equation (\ref{nishimori:pq-derivative}) then follows from 
(\ref{nishimori:pmq-relation1}).

\begin{figure}[t]
\begin{center}
\epsfigure{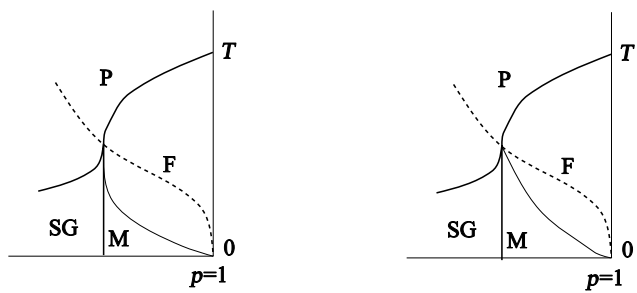}{3.5in}
\end{center}
\caption{Phase diagrams compatible (left) and incompatible (right)
with the relations proved in the text when the mixed phase
is characterized by a continuous part in $P_q(x)$.
}
\label{nishimori:fig3}
\end{figure}

The relation (\ref{nishimori:pq-derivative}) shows that the
distribution function of the spin glass order parameter remains
vanishing at almost all $x$ if the parameters $K$ and $p$ deviate
infinitesimally from their values satisfying $K=K_p$.
Thus the structure of $P_q(x;K)$ does not develop a continous
part under an infinitesimal deviation from the Nishimori line.
An interesting consequence is that the mixed phase with a continuous
part in $P_q(x)$, if it starts
to exist at the multicritical point, does not have a finite
extension under a very small deviation from the multicritical point.
Stated otherwise, the phase boundaries (the AT line and the boundary
between the mixed and spin glass phases) should merge
smoothly as they approach the multicritical point
as depicted on the left part of Fig. \ref{nishimori:fig3}.
A natural generic situation would be that both of these boundaries
are vertical
around the multicritical point as in the case of the SK model.
It should be noted that this argument does not apply when the RSB
in the mixed phase is characterized by several delta functions,
in which case the boundary with the structure
like the one on the right part of Fig. \ref{nishimori:fig3}
does not violate (\ref{nishimori:pq-derivative}).
%
\subsection*{Outline of the proof}

The proof of the relation (\ref{nishimori:pmq-relation1}) is not
very difficult \cite{nishimori:SGN}.
We start from writing out the configurational average of the
distribution function of magnetization
 \begin{equation}
   P_m(x; K)= \frac{1}{(2\cosh K_p)^{N_B}}
     \sum_{\{ \tau_{ij}=\pm 1\} } e^{K_p \sum \tau_{ij}}
    \frac{\sum_S \delta (x-N^{-1} \sum_i S_i )
     e^{K\sum \tau_{ij} S_i S_j}}
     {\sum_S e^{K\sum \tau_{ij} S_i S_j}}.
  \label{nishimori:Pm2}
 \end{equation}
The weight of configurational average ($p$ or $1-p$ for each bond)
is taken care of by
the factor $e^{K_p \sum \tau_{ij}}$ \cite{nishimori:hn81,nishimori:hn2001}.
The gauge transformation (\ref{nishimori:g-trsf}) changes the
above equation into the form
 \begin{equation}
   P_m(x; K)= \frac{1}{(2\cosh K_p)^{N_B}}
     \sum_{\{ \tau_{ij}=\pm 1\} } e^{K_p \sum \tau_{ij}\sigma_i \sigma_j}
    \frac{\sum_S \delta (x-N^{-1} \sum_i S_i \sigma_i)
     e^{K\sum \tau_{ij} S_i S_j}}
     {\sum_S e^{K\sum \tau_{ij} S_i S_j}}.
  \label{nishimori:Pm3}
 \end{equation}
It is relatively straightforward to see that the right-hand side
is equal to $P_q(x; K_p,K)$ by summing it up
over all possible $\{\sigma_i\}$, dividing the result by $2^N$,
and finally inserting an identity
$
   1=\sum_\sigma  e^{K_p \sum \tau_{ij}\sigma_i \sigma_j}/
   \sum_\sigma  e^{K_p \sum \tau_{ij}\sigma_i \sigma_j}
$
after the summation symbol over $\{\tau_{ij}\}$.

\section*{Conclusion}

We have proved that the AT line, if any, should lie below
the Nishimori line in the phase diagram of the
$\pm J$ Ising model on an arbitrary lattice.
It has also been argued that the mixed phase with a continous
part in the distribution function of the spin glass order
parameter, if any again,
does not have a finite extension around the multicritical point.
These results have been derived by a simple application of
gauge transformation.

The present work was supported by the Anglo-Japanese Collaboration
Programme between the Japan Society for the Promotion of Science
and The Royal Society.


\end{document}